\documentclass[aip,jcp,preprint,noshowkeys,superscriptaddress]{revtex4-2}

\usepackage{graphicx,bm,xcolor,microtype,multirow,amscd,amsmath,amssymb,amsfonts,physics,longtable,wrapfig,txfonts,soul}

\usepackage[utf8]{inputenc}
\usepackage[T1]{fontenc}
\usepackage{txfonts}

\usepackage[
	colorlinks=true,
    citecolor=blue,
    breaklinks=true
	]{hyperref}
\DeclareMathOperator*{\argmin}{argmin} 
\newcommand{\pinv}[1]{{#1}^{\oplus}}
\usepackage{scalerel}
\newcommand\Dag{\stretchrel*{\dag}{X}}

\begin{document}	


\title{ Many-body  $GW$  calculations with very large scale polarizable environments made affordable: a fully \textit{ab initio}  QM/QM  approach  }

\author{David Amblard}
\affiliation{Univ. Grenoble Alpes, CNRS, Inst NEEL, F-38042 Grenoble, France}
\author{Xavier Blase}
\affiliation{Univ. Grenoble Alpes, CNRS, Inst NEEL, F-38042 Grenoble, France}
\author{Ivan Duchemin}
\affiliation{Univ. Grenoble Alpes, CEA, IRIG-MEM-L\_Sim, 38054 Grenoble, France}
\email{ivan.duchemin@cea.fr}

\date{\today}

\begin{abstract}

We present  a many-body $GW$  formalism for quantum subsystems embedded in  discrete polarizable environments containing up to several hundred thousand atoms described at a fully \textit{ab initio} random phase approximation level.
Our approach is based on a fragment approximation in the construction of the Green's function and independent-electron susceptibilities.  
Further, the environing fragments susceptibility matrices  are reduced  to a minimal but accurate representation preserving  low order polarizability tensors through a constrained minimization scheme. 
This approach dramatically reduces the cost associated with inverting the Dyson equation for the screened Coulomb potential $W$, 
while preserving  the description of short to long-range screening effects. 
The efficiency and accuracy of the present   scheme is exemplified in the paradigmatic cases of 
fullerene bulk, surface, subsurface, and slabs with varying number of layers.  
\end{abstract}


\maketitle

\section{Introduction}
\label{sec:intro}

The description of the electronic  properties of  a quantum subsystem  embedded  in a  polarizable, or dielectric, environment (a molecular interface, a solvant, an electrode, etc.) remains a central issue in many fields pertaining to solid-state physics, chemistry or biology. Starting from the historical image charge models for electronic distributions close to a metallic surface  or within a dielectric cavity, \cite{Born_1920,Jackson} the need to describe the response of a polarizable environment to a charged (photoemission) or neutral (optical) excitation in a specific subsystem is still triggering significant developments to combine accuracy with numerical efficiency. In particular, the stabilization of an added hole or electron by the induced electronic rearrangements in a surrounding polarizable environment can be as large as several electronvolts. This so-called polarization energy, together with the additional effects of the electrostatic environment and wavefunction delocalization, strongly renormalize  the electronic properties.  In many situations, the environment is a complex, potentially infinite, system  that cannot be fully described at  the same quantum level as the subsystem of interest. 

Conceptually close to the historical models of image charges, the polarizable continuum model (PCM)   \cite{Miertus_1981,Cances_1997}   considers a quantum subsystem   located in a   cavity carved into a  medium described by an homogeneous macroscopic dielectric constant. 
As a more expensive alternative, discrete polarizable models, where atoms are described as  polarizable centers, allow for a more realistic description of screening inhomogeneities at short range in response to an electronic excitation in the quantum subsystem.
\cite{Thompson1995,Osted2006,Lin2007,Curutchet2009,Loco2021} 
Polarization energies converging slowly with environment size,  thousands of polarizable centers may be needed  in order to enter a regime where long-range extrapolation can be achieved on the basis of the calculated values.  This comes as a challenge to fully \textit{ab initio} approaches, triggering in practice the description of the environment at a semiclassical empirical level. In such semi-empirical discrete approaches, labeled generically QM/MM, or QM/MMpol to emphasize the polarizable nature of the environment,  atoms are provided with  effective polarizabilities that reproduce the correct molecular polarizability tensor and/or the macroscopic dielectric tensor of the material. \cite{D_Avino_2014,D_Avino_2016}     

Concerning the quantum mechanical formalism used to describe electronic excitations in the central subsystem, 
a specific class of   many-body perturbation theories, the $GW$ \cite{Hed65,Str80,Hyb86,God88,Lin88}
 and Bethe-Salpeter equation (BSE) \cite{Czanak_1971,Strinati_1988,Albrecht_1998,Rohlfing_2000,shirley-prl-1998} formalisms for the study of charged and neutral electronic excitations, have been recently combined with polarizable models of environment, both at the continuum \cite{Duchemin_2016,Duchemin_2018,Clary2023} and discrete \cite{Baumeier_2014,Li_2016,Li_2018,Wehner_2018} levels. 
 Indeed, while recent studies demonstrated that $GW$ calculations with cubic or even lower scaling could be achieved,
\cite{Rojas1995,Foerster2011,Neuhauser2014,Liu2016,Vlcek2017,Vlcek2018,Wilhelm2018,Forster2020,Kim2020,Kupetov2020,Duchemin2021,Wilhelm2021} 
the slow convergence of electrostatic and dielectric (screening) effects with respect to system size forbids a brute force treatment of complex environments within such approaches. 

The cost associated with the building of the   irreducible  electronic susceptibility is usually the bottleneck in 
time-dependent density-functional theory (TD-DFT) and $GW$ calculations. As a cure, fragments  approximations can dramatically reduce such a cost by neglecting wavefunction overlaps between weakly interacting subsystems. These fragment  or subsystem approaches have been recently implemented at the $GW$ and BSE levels in the case of systems presenting weakly interacting subunits, including molecular systems, \cite{Fujita_2018,Fujita_2019,Fujita_2021,Tolle_2021,Vlcek2021,Vlcek2023} interfaces,\cite{Neaton2006,Liu2019,Liu2020,Xuan2019} 2D materials, \cite{Andersen_2015,Winther_2017} but also in the less obvious case of covalent 2D systems.\cite{Amblard_2022} It remains that obtaining the screened Coulomb potential $W$ from the irreducible susceptibility, requiring a matrix inversion, restricts the number of fragments that can be dealt with at the fully \textit{ab initio} level. As such, the largest  fragment-based $GW$ calculations were obtained for a system containing about 800 benzene molecules (4800 non-H atoms)  in the case of molecular systems, an already remarkable achievement. \cite{Fujita_2021}  Similarly, we recently used a fragment $GW$ approach to partition a multilayer \textit{h}-BN system in up to 259 \textit{h}-BN fragments containing 66 non-hydrogen atoms each. \cite{Amblard_2022} 
The limiting factor to the simulation of larger systems was then the inversion of the Dyson equation to obtain $W$.
 
 In the present study, we introduce and assess  a fully  \textit{ab initio} scheme for embedded $GW$  calculations with hundreds of thousand atoms in the environment. Besides adopting the fragment approximation,  we search for an efficient low-rank representation of the susceptibility matrix
 associated with each fragment,   projecting them on-the-fly onto a minimal polarization basis preserving the dipolar, quadrupolar, etc. fragments polarizability tensors. As a result, the size of the Dyson equation for $W$ is dramatically reduced, while preserving the accuracy for the polarization energies at the meV level. We  explore the trade-off between accuracy and efficiency in the case of a fullerene crystal, both in the bulk, surface, subsurface and few-layers-slab limits.  
 

\section{ Theory}\label{Theory}

We very briefly outline the $GW$   formalism,   directing the reader to thorough reviews and books for a more detailed account on the subject.  \cite{Ary98,Farid99,Oni02,ReiningBook,Pin13,Gol19rev} 
We further introduce our embedding scheme associated with the definition of the environmental screening, or reaction field. 
Finally, we describe our   fitting   scheme allowing to dramatically reduce the size of the fragments dielectric matrix expressed in an effective polarization basis that preserves short to long-range screening effects. 

\subsection{ The $GW$  formalism }

Departing from the use of the electronic density  in DFT, the $GW$   formalism  takes as a central variable the  time-ordered one-body Green's function  built from  input $\lbrace \varepsilon_n, \phi_n \rbrace$  Kohn-Sham eigenstates, namely: 
 \begin{align}
 G({\bf r},{\bf r}'; \omega ) =  \sum_{n}    \frac{   \phi_n({\bf r}) \, \phi_n^*( {\bf r}')  }{ {\omega} - \varepsilon_n +i \eta \times \text{sgn}(\varepsilon_n -E_F) }  
 \label{GreenFunc}
 \end{align}
 where $\eta$ is a positive infinitesimal and $E_F$ the Fermi energy. Relying on perturbation theory to low order in the screened Coulomb potential $W$, the energy-dependent exchange-correlation self-energy $\Sigma({\bf r},{\bf r}' ; E)$ can be approximated by the $GW$ operator under the form: 
 \begin{align}
\Sigma^{GW}({\bf r},{\bf r}' ; E) = \frac{i}{2 \pi} \int_{-\infty}^{+\infty} \dd\omega \;  e^{i  \eta \omega  } \, G({\bf r},{\bf r}' ; E+ \omega)\,{W}({\bf r},{\bf r}' ;  \omega)
\label{sigmaGW}
\end{align}
 with $v$ the bare Coulomb potential and $W$ the dynamically screened Coulomb potential built within the random phase approximation (RPA): 
 \begin{align}
   {W}({\bf r},{\bf r}' ;  \omega) & = {v}({\bf r},{\bf r}' ) \nonumber \\
                  + &\int \dd{\bf r}_1 \dd{\bf r}_2   {v}({\bf r},{\bf r}_1 ) \;   \chi_0({\bf r}_1 ,{\bf r}_2 ;  \omega)  \, {W}({\bf r}_2,{\bf r}' ;  \omega) . \label{eqn:Dyson} 
 \end{align}
 Such an equation adopts a self-consistent Dyson-like form that needs to be inverted once the independent-electron   susceptibility has been built from Kohn-Sham one-body eigenstates:
\begin{align}
     \chi_0({\bf r},{\bf r}' ; \omega) & = \nonumber\\
       \sum_{m,n} &(f_m - f_n)  \frac{   \phi_m^*({\bf r})  \,\phi_n({\bf r})  \,      \phi_m({\bf r}') \, \phi_n^*({\bf r}')   }{ \omega - ( \varepsilon_n - \varepsilon_m) + i\eta\times\text{sgn}(\varepsilon_{n}-\varepsilon_{m}  )}  
    \label{eqn:xi0}
\end{align}    
 with $\lbrace f_{m/n} \rbrace$ level occupation numbers. The cost of calculating this independent-electron susceptibility grows as $\mathcal{O}(N^4)$ with respect to the number of electrons $N$ in the system.
Equivalently, dropping   the space variables for compactness, the Dyson equation can be formulated as:
 \begin{align}
 {W}( \omega)  = {v}  +    {v}  \,  \chi( \omega) \,  {v}, 
   \label{eqn:W_via_Chi}  
 \end{align}
 with $\chi$ the RPA  interacting susceptibility:
     \begin{align} 
   {\chi}( \omega) &= \chi_0(  \omega) +  \chi_0( \omega) \, {v}  \,\chi( \omega).         \label{eqn:xidyson}  
\end{align}
 The knowledge of $\Sigma^{GW}$ allows to correct the Kohn-Sham eigenvalues, replacing the density-based exchange-correlation potential $v^{XC}$ by the $GW$ self-energy:
 $$
 \varepsilon_n^{GW} = \varepsilon_n^{KS} + \langle \phi_n | \,\Sigma^{GW}( \varepsilon_n^{GW} )  -v^{XC} \,| \phi_n \rangle
 $$
 with $ \varepsilon_n^{GW} $ the so-called quasiparticle energies at which the self-energy operator must be calculated. 
 
\begin{figure}[t]
	\includegraphics[width=8.6cm]{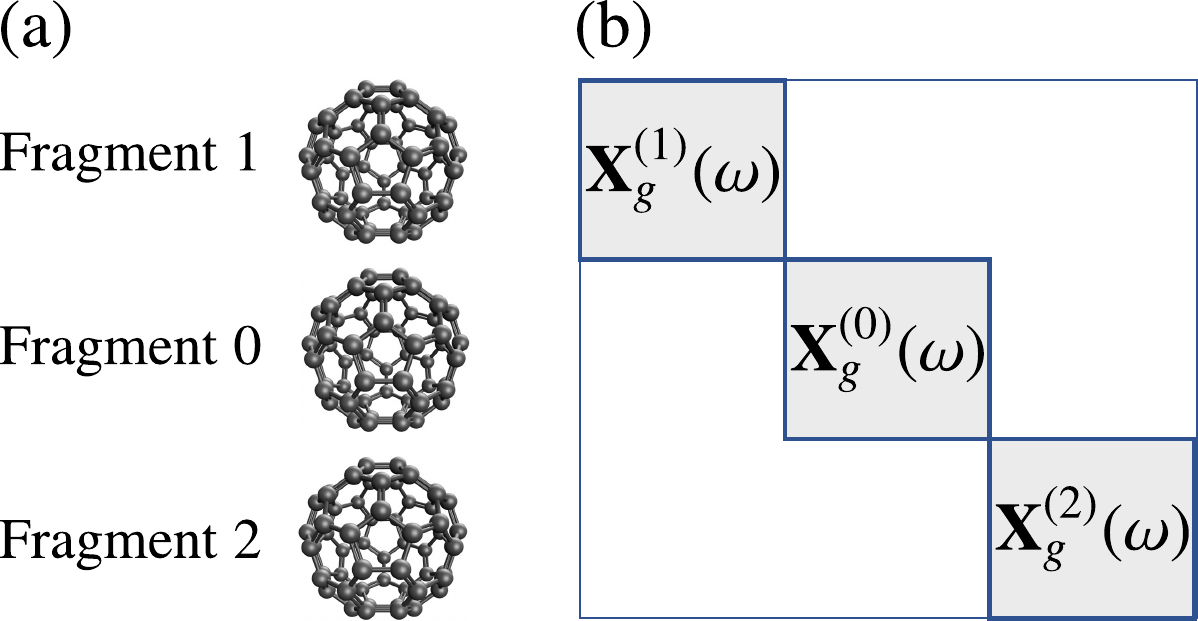}
	\caption{  Symbolic representation of the fragment approximation. Here the diagonal blocks are the interacting (reducible) susceptibility associated with the isolated fragments (in gas phase). As a result, the Coulomb potential in the associated Dyson equation should be the off-diagonal  $V^{(\mathrm{I} \ne \mathrm{J})}$ coupling only different blocks (see equation~\ref{eqn:xidyson-bis}). }
	\label{fig:scheme1}
\end{figure} 

\subsection{ Fragment approximation}

Our implementation \cite{Duchemin_2020,Duchemin_2021} of the $GW$  approach adopts a resolution-of-the-identity (RI) formalism \cite{Vahtras_1993,Ren_2012,Duchemin_2017b} where the density and its variations are expressed over a Gaussian auxiliary basis set $\lbrace P \rbrace$. The auxiliary basis functions must thus approximate the space generated by the products of molecular orbitals (MO) $\lbrace \phi_n \rbrace$:
$$
\phi_{n}({\bf r}) \,\phi_{m}({\bf r}) \overset{RI}{\simeq} \sum_{P} \mathcal{F}_{P}(\phi_n \phi_m)\,P({\bf r})
$$
leading in particular to fitting coefficients $\mathcal{F}_{P}(\phi_n \phi_m)=0$ when the $\phi_n$ and $\phi_m$  molecular orbitals are non-overlapping.  
Within this representation, Eqn.~\ref{eqn:xi0} rewrites:
\begin{equation}
     \chi_0({\bf r},{\bf r}' ; \omega)  
     \overset{RI}{\simeq}   \sum_{P,Q} {{X}}_0(P,Q\,;\omega)\, P({\bf r}) \,Q({\bf r}') 
    \label{eqn:xi0_1}
\end{equation}
with coefficients 
\begin{align}
{X}_0(P,Q\,;\omega)  =  
       \sum_{m,n} &(f_m - f_n)  \frac{   \mathcal{F}_{P}(\phi_m^*\phi_n)    \;  \mathcal{F}_{Q}(\phi_m\phi_n^*)  }{ \omega - ( \varepsilon_n - \varepsilon_m) + i\eta\times\text{sgn}(\varepsilon_{n}-\varepsilon_{m}  )}   \;.
    \label{eqn:xi0_2}
\end{align}

In the fragment approximation, the full system auxiliary basis is simply the union of the subsystem basis sets, while each subsystem density can be expressed in its own corresponding basis. 
In such a case, the analysis of Eqn.~\ref{eqn:xi0_2} indicates that the joint contribution of two non overlapping subsystems to the representation of the independent-electron susceptibility $\chi_0({\bf r},{\bf r}' ; \omega) $ should be zero. 
Within that limit, the RI-fitted independent-electron susceptibility matrix is thus block diagonal, with blocks corresponding to the constituting subsystems gas phase (isolated) irreducible susceptibilities.  

Labelling $ {\bf X}_0^{(\mathrm{I})}$ the block of fit coefficients corresponding to the independent-electron susceptibility of subsystem (I), the Dyson equation for the full system interacting susceptibility coefficients  matrix ${\bf X}(\omega)$ (eqn.~\ref{eqn:xidyson}) reads:
\begin{align}
 {\bf X}(\omega)^{-1} &= \sum_\mathrm{I} {\bf X}_0^{(\mathrm{I})} (\omega)^{-1} - \sum_{\mathrm{I,J}} {\bf V}^{(\mathrm{IJ})}    
\end{align}
with ${\bf V}^{(\mathrm{IJ})}$ the block corresponding to the Coulomb interactions between auxiliary basis elements of fragments (I) and (J). This equation  can also be conveniently rewritten:
%
\begin{align}
 {\bf X}(\omega)^{-1} &= \sum_\mathrm{I} {\bf X}_g^{(\mathrm{I})} (\omega)^{-1} - \sum_{\mathrm{I\neq J}} {\bf V}^{(\mathrm{IJ})}    \label{eqn:xidyson-bis}
\end{align}
with ${\bf X}_{g}^{(\mathrm{I})}(\omega)$ the block of coefficients for the isolated (gas phase) interacting susceptibility of fragment (I). In this latter formulation, only the off-diagonal Coulomb interactions ${\bf V}^{(\mathrm{I\neq J})} $ that accounts for inter-fragments coupling are considered (see Fig.~\ref{fig:scheme1}). Similar equations can be found in the framework of subsystem TD-DFT. \cite{Pavanello2013,Tolle2022}

In the fragment approximation, the cost of calculating all fragments ${\bf X}_0^{(\mathrm{I})}$ and ${\bf X}^{(\mathrm{I})}_{g} $  blocks scales linearly   with respect to the number of fragments. 
This is a considerable saving and it is now the inversion of the Dyson equation (Eqns.~\ref{eqn:Dyson} or \ref{eqn:xidyson-bis}), with  cubic scaling with respect to the total number of fragments, that becomes the bottleneck in the limit of a very large number of subsystems. It is such a problem that we address below  by optimally reducing the size of each ${\bf X}_{g}^{(\mathrm{I})}$  susceptibility representation.

\subsection{ Constrained reduction of the  fragment susceptibilities }\label{cstr_reduc}

 A typical calculation that expands MOs over a triple-zeta def2-TZVP basis set \cite{weigend2005a} involves the  corresponding optimized auxiliary def2-TZVP-RI basis set \cite{weigend1998a} that is composed of 95 orbitals for e.g. B, C, N, O atoms. In this situation, an environment containing of the order of 10$^5$ atoms will result in  susceptibility matrices of the order of 10$^7$ in size to be dealt with in the Dyson equation.

For the fragment of interest, for which we want to actually perform a $GW$ correction, we  preserve the full auxiliary basis optimized for the corresponding MO basis sets. However, concerning the fragments in the environment, we emphasize that we are mainly interested in their contribution to the reaction field, namely to the induced dipoles, quadrupoles, etc., developed as a response to an electronic excitation in the central subsystem. As such, the full details of the susceptibility in the auxiliary $\lbrace P \rbrace$ basis may not be necessary. 

In order to reduce the computational effort of the Dyson equation, we therefore look for an efficient and compact way to represent the interacting susceptibility of the fragments in the environment. We seek a lower-rank approximation to the gas phase interacting susceptibilities $\chi^{(\mathrm{I})}_{g}(\omega)$ :
\begin{align}
{\chi}^{(\mathrm{I})}_g(\bf{r},\bf{r'};\omega) & \;\;\overset{RI}{\simeq} & \!\!\!\!\!\sum_{P,Q} X_g^{(\mathrm{I})}(P,Q\,;\omega) \;  P({\bf{r} }) \, Q({\bf{r'}}) \label{eqn:xiref} \\
& \!\!\!\overset{MODEL}{\simeq} & \!\!\!\!\! \sum_{\gamma,\gamma'} \widetilde{X}_g^{(\mathrm{I})}(\gamma,\gamma'\,;\omega) \; \gamma({\bf{r} }) \, \gamma'({\bf{r'}})
 \label{eqn:ximod_def}
\end{align}
where we use a small basis sets $\lbrace \gamma \rbrace$, which could be for example a minimal Gaussian $(\textit{sp}^3$) 4-orbitals basis per atom in order to mimic the  induced charges-and-dipoles models developed in QM/MM techniques.\cite{D_Avino_2014}   

For the isolated fragment (I), the resulting  errors in the interacting susceptibility and its corresponding contribution to the screening, or reaction,   field 
respectively read:
\begin{equation}
\begin{split}
\Delta \chi_{g}^{(\mathrm{I})}({\bf r},{ \bf r'};\omega) & =
\sum_{\gamma,\gamma'} \widetilde{X}_g^{(\mathrm{I})}(\gamma,\gamma'\,;\omega) \; \gamma({\bf{r} }) \, \gamma'({\bf{r'}}) \\
& -\sum_{P,Q} X_g^{(\mathrm{I})}(P,Q\,;\omega) \;  P({\bf{r} }) \, Q({\bf{r'}})
 \label{eqn:dximod}
\end{split}
\end{equation}
and
\begin{equation}
\begin{split}
& \Delta v_{\text{screen}}^{(\mathrm{I})}({\bf r},{\bf r'};\omega) \phantom{\Bigg)}\\
& =  \iint \dd{\bf r}_1 \dd{\bf r}_2 \; v({\bf r},{\bf r}_1)\  
\Delta \chi_{g}^{(\mathrm{I})}({\bf r}_1,{\bf r}_2;\omega)\  v({\bf r}_2,{\bf r}').
 \label{eqn:dvreac}
\end{split}
\end{equation}
Once the model ``polarization'' basis $\lbrace \gamma \rbrace$ is fixed, the associated $\widetilde{\mathbf{X}}_g^{(\mathrm{I})}$ set of coefficients can thus be simply deduced by minimizing the error in the screening field $\Delta v_{\text{screen}}^{(\mathrm{I})}$ through a set $\{t\}$ of test functions:
\begin{equation}
\begin{split}
\widetilde{\mathbf{X}}_g^{(\mathrm{I})}(\omega) = \argmin_{\big\{\widetilde{X}_g^{(\mathrm{I})}(\gamma,\gamma'\,;\omega)\big\}} 
\sum_{t,t'} \left| \langle t\, | \Delta v_{\text{screen}}^{(\mathrm{I})}(\omega) |\, t' \rangle \right|^2.  
 \label{eqn:ximod_coeffs}
\end{split}
\end{equation}
The salient features of this equation is that, since we measure the difference in the reaction field on a set of test functions, we are free to focus on the specific components of the screening field $v \chi v$ that we want to preserve. Details about the resolution of this equation are given in Appendix \ref{appendix_coeff}.

\begin{figure}[t]
	\includegraphics[width=8.6cm]{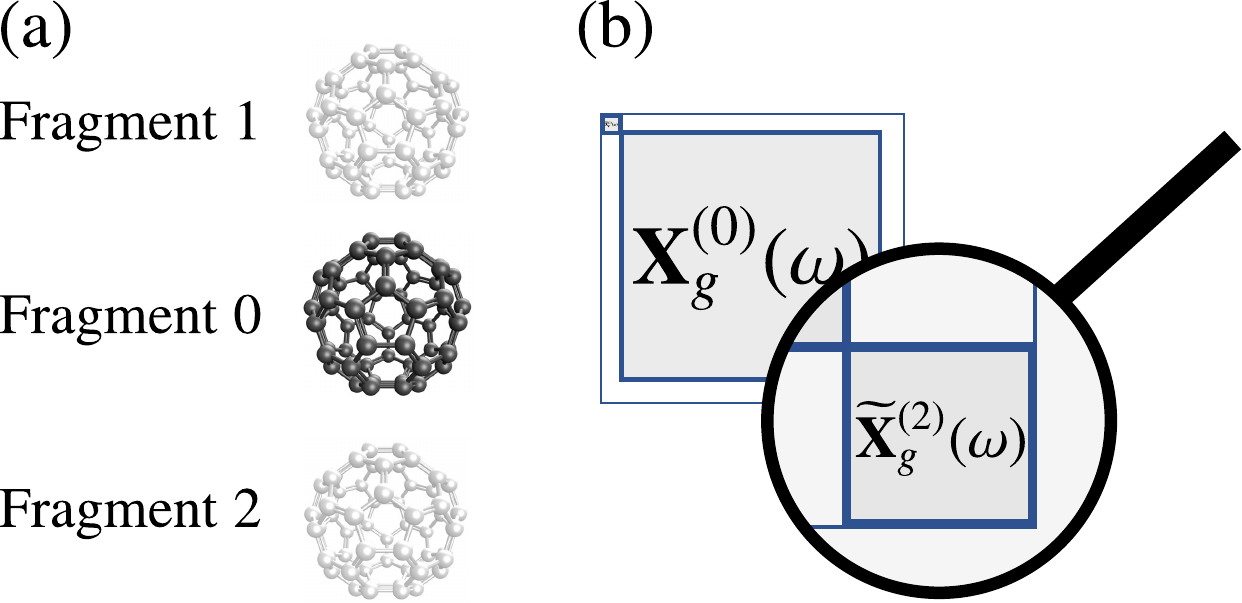}
	\caption{  Sketch of the  susceptibility blocks compression   associated with the fragments in the environment (shaded $C_{60}$). The susceptibility of the central fragment of interest (fragment I=0), on which will be performed the embedded $GW$ calculation, is not reduced.}
	\label{fig:scheme3}
\end{figure}

A naive choice for the test functions $\lbrace t \rbrace$ could be the set of auxiliary functions $\lbrace P \rbrace$ used to construct the reference $\chi^{(\mathrm{I})}_g$ response function. 
As shown below,  this strategy is rather inefficient. The reason for this failure is that we do not intend to use the model susceptibility to perform a $GW$ calculation on fragment (I) itself.  Instead, we need it to build the fully interacting screening potential $v\chi v$ through the Dyson equation (Eq.~\ref{eqn:xidyson-bis}) that couples $(\mathrm{I}\ne \mathrm{J})$ fragments,  focusing in the end on the central fragment ($\mathrm{I}=0$ in Fig.~\ref{fig:scheme3}) on which we perform the $GW$ calculation.
As such, priority should be given to the interactions between the different model fragments, starting from neighboring fragments up to the long range interactions dominated by low order momenta of their polarizability tensors. 

A very simple yet successful strategy consists in keeping the test functions localized on the atoms of the fragment (I) for which we seek the model susceptibility, but building the test set $\lbrace t  \rbrace$ out of very diffuse orbitals that will sample the surrounding fragments. The basic idea is that using such diffuse test functions allows to ``reach out'' for the effect of $v \chi^{(\mathrm{I})} v$ on neighboring molecules. The test set can then be completed with the auxiliary basis $\lbrace P \rbrace$ associated with fragment (I), but down-weighted, in order to keep the emphasis on the diffuse orbitals during the minimization process. More details about this test basis can be found in Appendix \ref{annexe:test_basis}.

Simultaneously, the preservation of long range interactions can be guaranteed by enforcing low order Cartesian momenta of the susceptibility through Lagrange multipliers:
\begin{align}
    \left\langle  x^{m}\, y^{n}\,  z^{p}  \left|  \; {\Delta \chi}^{(\mathrm{I})}_g(\omega) \right|  x^{m'} \, y^{n'} \, z^{p'}   \right\rangle  = 0.
 \label{eqn:lagrange}
\end{align}
In the following, we label $l_{max}$ the maximum order $m+n+p$ enforced for a specific model susceptibility. For example, $l_{max}=1$ corresponds to the preservation of the fragment neutral monopole, as well as the dipolar polarizability tensor. Imposing such a constraint, along with the use of diffuse test functions, ensures that  the reaction field  will be well reproduced not only in the vicinity of fragment (I), but in the long-range as well. Imposing higher order polarizability tensors momenta (e.g. $l_{max}=2$), can be achieved but with a mild impact as discussed below.



\subsection{ Minimal effective polarizability basis }

So far the choice of the $\lbrace \gamma \rbrace$ has been left arbitrary. 
Contrary to the induced charge-and-dipole models used in semi-empirical QM/MM techniques,
we follow here a more automated route. The polarization basis can be obtained as the result of a generalized minimization process, that is we include the $\lbrace \gamma \rbrace$ in the minimization process: 
\begin{equation}
\begin{split}
\argmin_{\{\gamma\} } \left(\;\min_{\big\{\widetilde{X}_g^{(\mathrm{I})}(\gamma,\gamma'\,;\omega)\big\}} 
\sum_{t,t'} \left| \langle t\, | \Delta v_{\text{screen}}^{(\mathrm{I})}(\omega) |\, t' \rangle \right|^2  \;\right)
 \label{eqn:ximod_dirs}
\end{split}
\end{equation}
where the only input choice is now the number $N_P$ of $\gamma$ polarization vectors. 
In practice, the $\gamma$ functions are  expressed in the $\lbrace P  \rbrace$ auxiliary basis set associated with fragment (I), namely
\begin{equation}\label{extraction_gamma}
\gamma({\bf{r}}) = \sum_{P} C_{\gamma P}\;  P({\bf{r}})\;,
\end{equation}
so that the $\lbrace C_{\gamma P} \rbrace$ coefficients   are now the minimization variables. We tackle this somewhat complex minimization problem by iterating over two distinct steps: i) inner optimization of the $\widetilde{\mathbf{X}}_g^{(\mathrm{I})}(\omega) $ matrix elements at fixed $\lbrace \gamma \rbrace$ (see equation~\ref{eqn:ximod_coeffs}), which is solved exactly through linear algebra; ii) outer optimization of the $\lbrace C_{\gamma P} \rbrace$ coefficients at fixed $\widetilde{\mathbf{X}}_g^{(\mathrm{I})}(\omega)$ matrix elements which is done using  gradient descent  techniques. 
More details about this last step can be found in Appendix \ref{annexe:pola_basis}.

In the simplest case where test functions $\lbrace t \rbrace$ only span the auxiliary $\lbrace P \rbrace$ set, and in the absence of constraint, the Eckart–Young–Mirsky theorem states that the   $N_P$   optimal $\lbrace \gamma \rbrace$ polarization vectors are similar to those defined in Refs.~\citenum{Wilson08,Govoni15}, namely  the leading eigenvectors of the so-called symmetrized susceptibility. 
The present minimization formulation allows further flexibility with the introduction of test functions and constraints, allowing on-the-fly design of model dielectric functions for specific purposes,  emphasizing  short-to-long-range or on-site accuracy.

 We conclude this Section by emphasizing again that the operations described above (calculations of the reference $\mathbf{X}_g^{(\mathrm{I})}(\omega)$ and $\widetilde{\mathbf{X}}_g^{(\mathrm{I})}(\omega)$ matrices, SVD decomposition of related operators, etc.) are performed on isolated fragments, leading to a computational cost that is linear in the number of distinct fragments. In turn, the number of operations related to inverting the Dyson equation for the total screened Coulomb potential, involving interactions between all fragments, is dramatically reduced through reduction of the associated prefactor. As shown below, the optimal polarization basis can be made typically 10$^2$  times smaller than the original auxiliary basis set, preserving the polarization energy in the meV range, leading to a reduction of the order of 10$^6$ of the cost associated with obtaining an accurate $W$ operator on the central fragment (I=0).



\subsection{ Technical details }
 
The present subsystem approach   with minimal representation of the fragments electronic susceptibility   has been implemented in the \textsc{beDeft} (beyondDFT) package. 
\cite{Duchemin_2020,Duchemin_2021} Input Kohn-Sham eigenstates are generated at the def2-TZVP   PBE0 \cite{perdew-bcp-1996,adamo-jcp-1999} level with the \textsc{Orca} package. \cite{orca,orca-5.0} 
We adopt the corresponding  def2-TZVP-RI  auxiliary basis sets associated with the Coulomb-fitting resolution-of-the-identity (RI-V) approach. \cite{Vahtras_1993,duchemin-jctc-2017}  
The molecular geometries for C$_{60}$  and the pentacene are obtained at the def2-TZVP PBE0 level. The face-centered cubic (fcc) C$_{60}$ dense phase is constructed taking experimental lattice parameters\cite{Heiney1991} (a = 14.17~{\AA}), neglecting orientational disorder. The C$_{60}$ surface we consider is the (111) surface. 

Even though the present scheme   allows to compute reaction fields at finite (imaginary) frequencies, following the  analytic continuation approach to the dynamical $GW$ self-energy implemented in \textsc{beDeft},\cite{Duchemin_2020} we calculate here the polarization energies at the static Coulomb-Hole plus Screened-Exchange (COHSEX) level,\cite{Hed65} with:
\begin{align}
\Sigma^{\text{SEX}}({\bf r},{\bf r}') &= - \sum_n^{\text{occp}}  \phi_n({\bf r}) \,\phi^{*}_n({\bf r}') \, W({\bf r},{\bf r}';\omega=0) \label{SEX}\\
\Sigma^{\text{COH}}({\bf r},{\bf r}') &= \frac{1}{2} \,\delta( {\bf r}-{\bf r}') \left[ W({\bf r},{\bf r}';\omega=0)-v({\bf r},{\bf r}') \right] \label{COH}
\end{align}
with the screened-exchange term involving a summation over occupied (occp) levels only. Following previous studies, \cite{Li_2016,Fujita_2018,Li_2018,Fujita_2019,Fujita_2021,Tolle_2021} our polarization energy $P_{n}$ for a given energy level is taken to be the difference between the static COHSEX energy level in the presence of a polarizable environment and its analog in the gas phase, namely: 
$$
P_n = \varepsilon_n^{GW_\text{e}} - \varepsilon_n^{GW_\text{g}} \simeq  \varepsilon_n^{\text{COHSEX}_\text{e}} - \varepsilon_n^{\text{COHSEX}_\text{g}}
$$
 where the index (e) and (g) in $GW_{\text{e}/\text{g}}$ and $\text{COHSEX}_{\text{e}/\text{g}}$ stand for embedded (e) and gas (g) phases. 
Such a definition is consistent with standard PCM  implementations where the macroscopic dielectric constant is taken to be the optical one in the low frequency limit.
Similarly, in standard QM/MM implementations, the semi-empirical atomic polarizabilities are designed to reproduce the fragment electronic polarizability in the static limit. \cite{D_Avino_2014}
Extension to dynamical reaction fields   will be discussed in subsequent studies.

From the knowledge of such a polarization energy $P_n$, calculated at the static $\Delta$COHSEX level,
the absolute quasiparticle energy can be obtained as:
\begin{equation*}
    \varepsilon_n^{GW_\text{e}} = \varepsilon_n^{GW_\text{g}} + P_n^{{\Delta}\text{COHSEX}}. 
\end{equation*}
In the fragment approximation, and in the absence of wavefunction hybridization, such a value yields the energy of the corresponding band center.\footnote{ In the fragment approximation, band dispersion originating from wavefunction hybridization between fragments cannot be accounted for.}
In particular, one can recover the experimental peak-to-peak gap in the dense phase, namely the difference of energy between the highest-occupied and lowest-unoccupied  molecular orbital  (HOMO/LUMO) band centers. 
In the following, when needed, the gas phase $\varepsilon_n^{GW_\text{g}}$ quasiparticle energy levels will be calculated at the partially self-consistent ev$GW$@PBE0 level, that has been shown \cite{Rangel2016,Kaplan2016}
to be more accurate than non-self-consistent calculations,
unless an optimally tuned functional is used for the starting DFT Kohn-Sham calculation. \cite{Bruneval2013,Rangel2016}

 Finally, we only need to perform an explicit $GW$ correction for the fragment of interest ($\mathrm{I}=0$). In other words, while the interacting susceptibility matrix $\mathbf{X} (\omega=0)$  of Eq.~\ref{eqn:xidyson-bis} is defined for the full system, the screened Coulomb potential matrix $\mathbf{W}=\mathbf{V}+\mathbf{V}\mathbf{X}\mathbf{V}$ entering  Eqs.~\ref{SEX} and \ref{COH} is only computed explicitly for the corrected fragment. This enables us to save both on memory footprint and CPU time aspects.

\section{Results}

  \subsection{ Validation }


We start by looking at the evolution of the static dipolar polarizability tensor for a given fullerene, in the gas phase, obtained with the model susceptibility matrix as a function of $N_P$ (the number of polarization vectors we keep). Namely, we compute 
\begin{equation} 
\left[{\alpha}\right]_{ij}  = -  \int \dd{\bf{r}} \dd{\bf{r}'} r_i \left( {\sum_{\gamma,\gamma'}^{N_P} \gamma({\bf{r}})   \widetilde{X}_g(\gamma,\gamma';\omega=0)  \gamma'({\bf{r'}})} \right) r'_j. \label{pola}
\end{equation}
This tensor is a key quantity for the long-range screening effects originating from a given fragment and represents thus a  direct measure of the accuracy of the fitted susceptibility. 
\begin{figure}[htbp]
	\includegraphics[width=8.6cm]{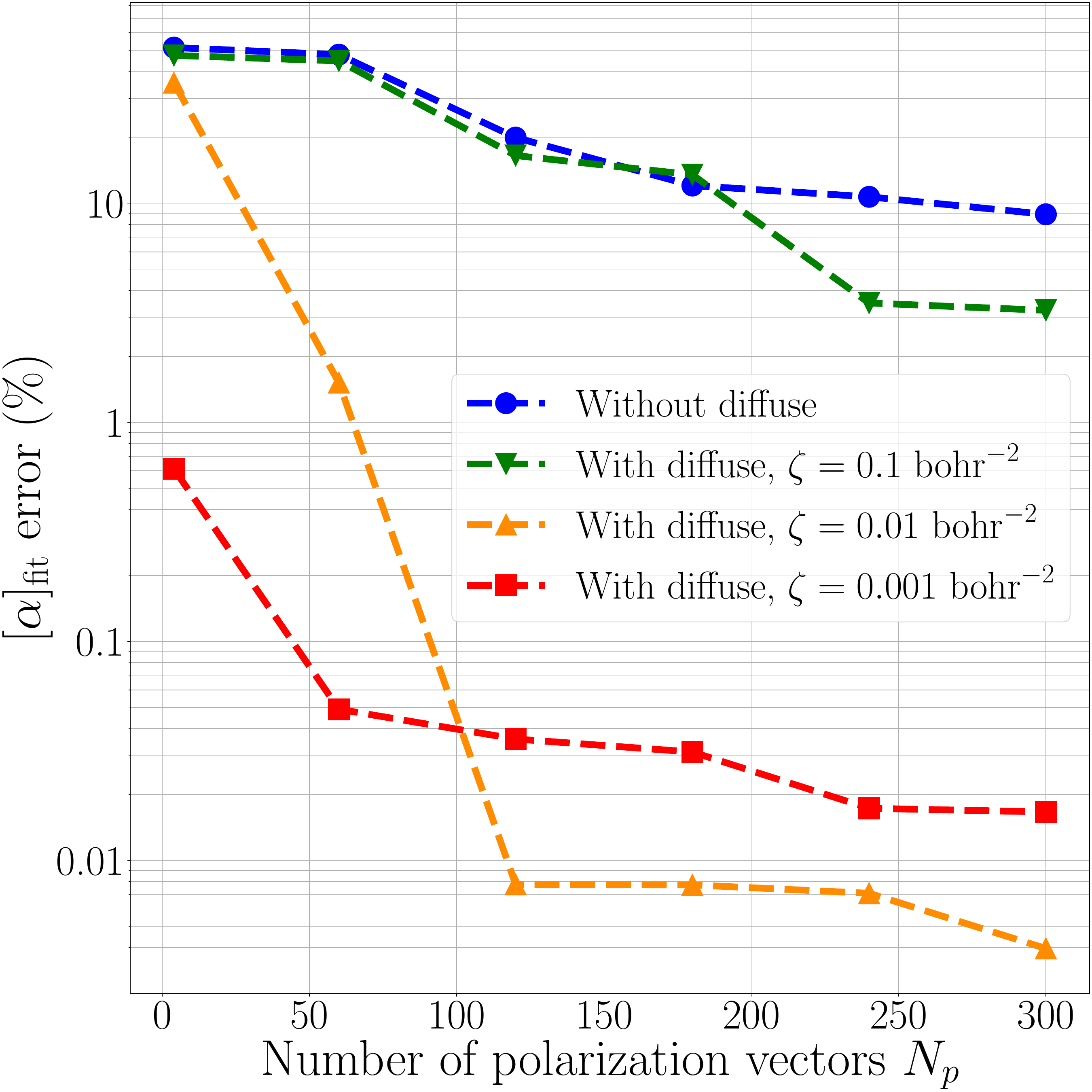}
	\caption{ Evolution of the relative error $\big\|[\alpha]_{\text{fit}}-[\alpha]_{\mathrm{ref}}\big\|\,/\,\big\|[\alpha]_{\mathrm{ref}}\big\|$ (in percentage) of a fullerene RPA dipolar polarizability tensor $[\alpha]$ (see Eq.~\ref{pola}) as a function of the number $N_P$ of polarization  vectors per C$_{60}$ (maximum number 5700). $||\cdot||$ corresponds to the Frobenius norm. $[\alpha]_{\text{ref}}$ is computed with the susceptibility described in the full auxiliary basis (with 5700 vectors), while $[\alpha]_{\text{fit}}$ is calculated using only $N_p$ polarization vectors. Results for different choices of   test functions are plotted.}
	\label{fig:polar}
\end{figure} 
First, we explore the strategy where the test functions $\lbrace t \rbrace$ are taken to span the auxiliary basis $\lbrace P \rbrace$ located on the fragment (a fullerene) for which we build the model susceptibility. Relative errors on the (Frobenius) norm of the dipolar polarizability tensor, with respect to a reference calculation using the full auxiliary basis (5700 vectors), are represented in Fig. \ref{fig:polar} (blue dots). As expected, this error decreases as the number of polarization vectors increases. For $N_P=240$, the relative error is of the order of $10\%$, that is still rather large. This number $N_P = 240$ corresponds to a typical minimal $\textit{sp}^3$ basis per atom of the kind used in semi-empirical induced charges-and-dipoles polarizable models. 

We now perform the same exercise but adding to the test functions $\lbrace t \rbrace$ a set of atom-centered diffuse Gaussian orbitals. Such diffuse functions are typically one set of ($\textit{s,p,d,f,g}$) orbitals per atom with, for sake of simplicity,  the same $e^{- \zeta r^2}$  radial part. Results for different values of $\zeta$ are reported on Fig.~\ref{fig:polar}. A value of $\zeta=0.1$ bohr$^{-2}$ (green down triangles), comparable to 0.2  bohr$^{-2}$  for the most diffuse carbon atomic orbital in the def2-TZVP-RI basis set, does not improve the quality of the fit.  Increasing the diffuse character of these functions, with $\zeta=0.01$ bohr$^{-2}$ (orange up triangles) improves significantly the quality of the results. For $N_{p}=120$, namely $\simeq 2\%$ of the dimension of the original auxiliary $\lbrace P \rbrace$ basis set, the relative error is below $0.01\%$. Increasing too much the extent of the diffuse orbitals, with e.g. $\zeta=0.001$ bohr$^{-2}$ (red squares) degrades the quality of the results. Even if relative errors are smaller with such a small $\zeta$ value in the limit of a very small number of polarization vectors ($N_{p}=4$ or $60$), $\zeta=0.001$ bohr$^{-2}$ leads to greater errors than  $\zeta=0.01$ bohr$^{-2}$ for larger values of $N_p$.

\begin{figure}[htbp]
	\includegraphics[width=8.6cm]{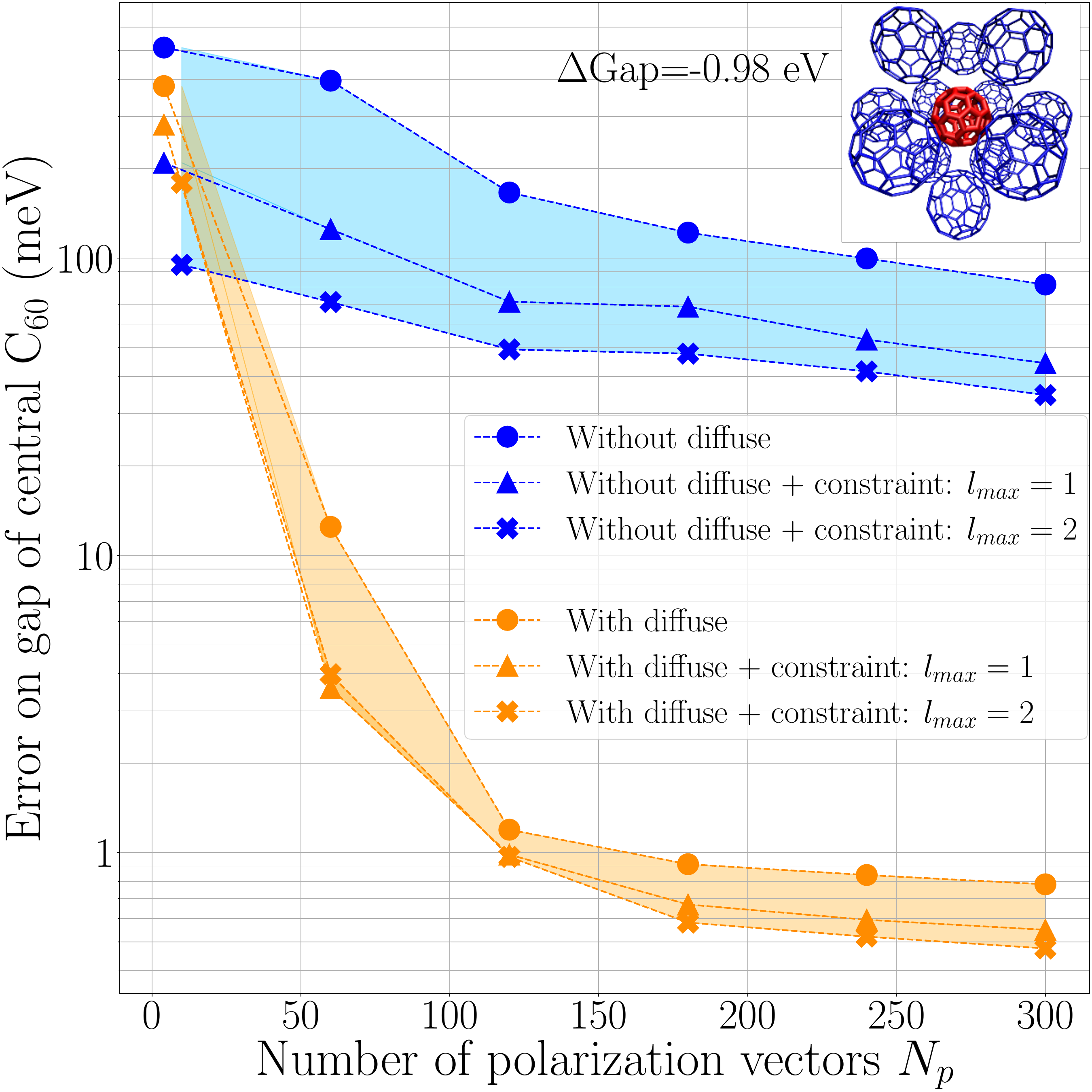}
	\caption{ Error on the central $C_{60}$ [red, Inset] gap  as a function of the number $N_P$ of polarization vectors per C$_{60}$ in the first-neighbors shell [blue, Inset].  Energies on the ordinates are in meV and log-scale. Results for test functions with and without diffuse orbitals, with and without constraints (see section \ref{cstr_reduc}), are shown. Diffuse functions use $\zeta=0.01$ bohr$^{-2}$. 
 The reference $\Delta\text{Gap}=-0.98$~eV corresponds to the gap reduction from the gas phase to the 13-C$_{60}$ cluster when all fragments are described by their full basis ($N_{p}=5700$).  
 }
	\label{fig:fig1}
\end{figure}

The quality of the polarizability tensor obtained with the low-rank susceptibility insures that long-range interactions will be accurately reproduced. We now focus on nearest-neighbor interactions. We study in particular the HOMO/LUMO energy gap associated with a fullerene (in red in Fig.~\ref{fig:fig1} Inset) surrounded by its first shell of 12 nearest-neighbors (in blue).


In a standard fragment calculation at the full def2-TZVP/def2-TZVP-RI level, the  central C$_{60}$ HOMO-LUMO gap  closes by $\sim$0.98 eV  due to the enhanced screening induced by the first shell of neighbors. This represents   about $60\%$ of the total polarization energy (see Section \ref{C60_env}) as compared to a fullerene in a fullerite, namely an infinite fullerene crystal. 

We now study the effect of reducing the size $N_P$ of the polarization basis on the 12 surrounding C$_{60}$. As previously, we start by using test functions taken only in the span of the auxiliary basis $\lbrace P \rbrace$ of the fragment (a fullerene) whose model susceptibility is fitted. 
The results are provided on Fig.~\ref{fig:fig1} (blue dots). As expected, the error on the central C$_{60}$ HOMO-LUMO gap, as compared  to the reference calculation, decreases with the number of polarization vectors. For $N_P=240$, the error is of the order of 100 meV, allowing to have a qualitative result but representing still an error of the order of 10$\%$ with respect to the targeted polarization energy. 

Similarly to the previous study of the polarizability, we now add diffuse functions in the test basis, with $\zeta=0.01$ bohr$^{-2}$. The related evolution of the  error on the polarization energy for the gap  is represented in Fig.~\ref{fig:fig1} (orange dots). Clearly, the addition of diffuse functions, allowing to test the quality of the model $\mathbf{V}  \widetilde{\mathbf{X}}_{g}^{(\mathrm{I})} \mathbf{V}$ reaction field in the vicinity of molecule (I),  dramatically accelerates the convergence of the polarization energy with respect to the size of the model susceptibility matrix. For $N_P = 180$, namely $\simeq 3\%$ of the  original susceptibility matrix size (5700) for one C$_{60}$, the error is now of the order of 1 meV, reaching  quantitative accuracy. 

We further add the constraints (equation~\ref{eqn:lagrange}), with and without diffuse functions, to enforce  the exact dipolar polarizability tensor with $l_{max}=1$ (up triangles in Fig.~{\ref{fig:fig1}}), or up to second order moments with $l_{max}=2$ (crosses in Fig.~{\ref{fig:fig1}).  In all cases, the constraints improve the accuracy, in particular in the small $N_P$ limit, even though their impact is not as important as adding diffuse test functions. Such a behaviour can be understood by looking, e.g., at Fig.~\ref{fig:polar} for $\zeta=0.01$ bohr$^{-2}$ and $N_P=120$. The dipolar polarizability is already quite well reproduced so that the constraint leads to a small improvement. Fig.~{\ref{fig:fig1}} reveals that the constraint $l_{max}=2$ improves very slightly the error on the gap, in comparison to the constraint $l_{max}=1$. When diffuse function are added to the test basis, the differences between $l_{max}=1$ (orange up triangles) and $l_{max}=2$ (oranges crosses) are less than $0.1$ meV for $N_{p}\ge 120$. Since imposing the constraint comes at no cost, we keep $l_{max}=1$ in the forthcoming calculations.


The test provided above for the polarization energy originating from the first-nearest-neighbors is the most stringent test. For fragments located farther away, the dipolar component of the reaction field, that we strictly impose, becomes more and more dominant. This is illustrated in Fig.~\ref{fig:fig_55C60} where we study the HOMO-LUMO gap of a C$_{60}$ surrounded now by its two nearest-neighbor shells (see Inset   Fig.~\ref{fig:fig_55C60}). The size of this cluster amounts to 55 fullerenes. When all fullerenes  are described at their full def2-TZVP/def2-TZVP-RI level (in the fragment approximation), the gap of the central fullerene closes by $1.25$ eV from the gas phase to the 55-C$_{60}$ cluster geometry.
We study the effect of reducing the size $N_p$ of the polarization basis used to describe the susceptibility of the 42 surrounding C$_{60}$ in the second shell, keeping the full auxiliary basis to describe the central fullerene and its first-nearest neighbors shell. As such, we mainly focus on the error induced by the fitting process on fragments located at middle to long-range of the central subsystem of interest.

\begin{figure}[htbp]
	\includegraphics[width=8.6cm]{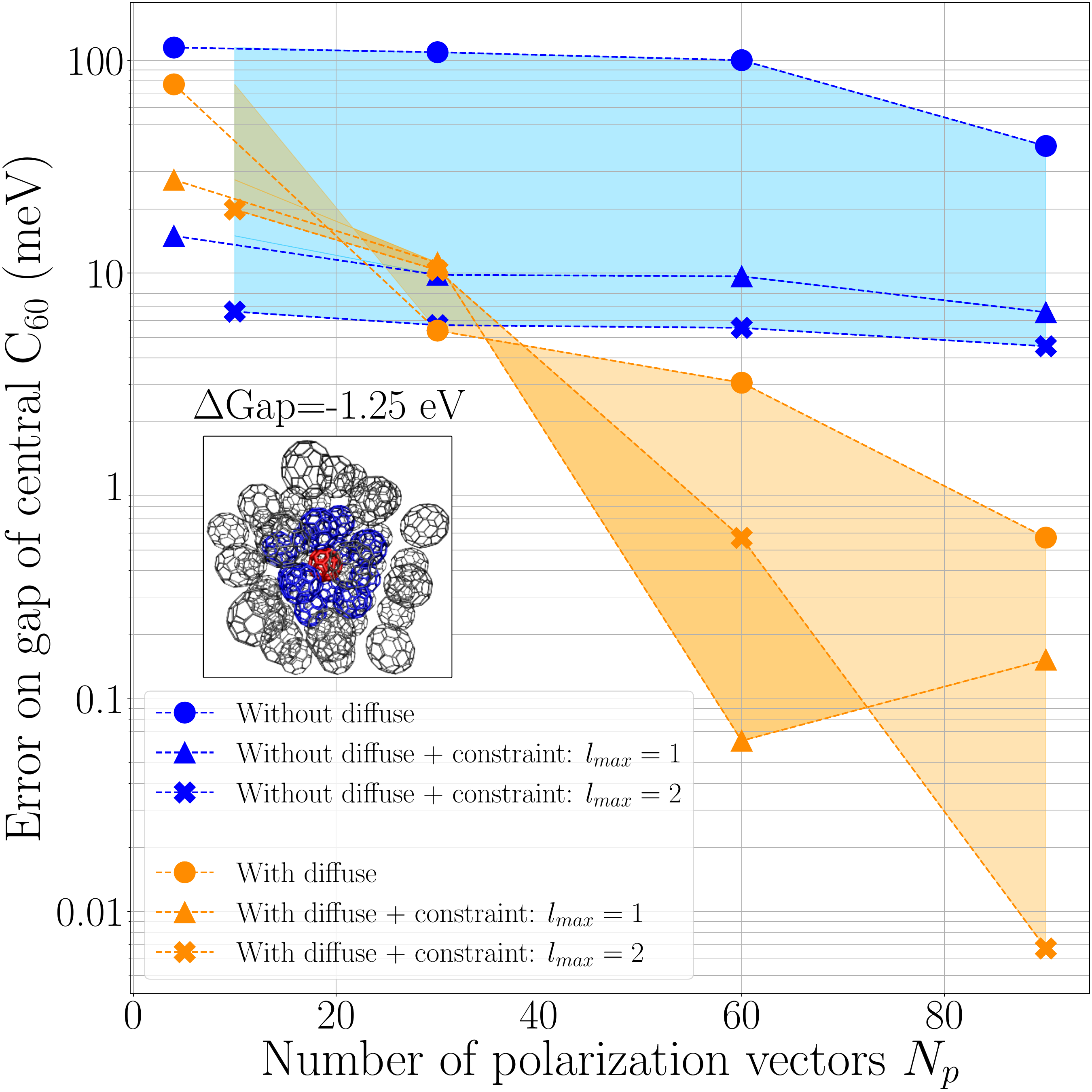}
	\caption{ Error on the gap for a fullerene surrounded by its two first-shells of neighbors. The susceptibility of the central (in red, Inset) and 12 first-nearest-neighbors (in blue, Inset) C$_{60}$  are described by the full auxiliary basis (5700  orbitals), while the susceptibility for each of the 42 C$_{60}$ in the second shell of neighbors (in grey, Inset) is described by $N_P$   polarization vectors. Energies on the ordinates are in meV and log-scale. Results for test functions with and without diffuse orbitals, with and without constraints (see section \ref{cstr_reduc}), are shown. Diffuse functions use $\zeta=0.01$ bohr$^{-2}$ (see text). 
 The value $\Delta\text{Gap}=-1.25$~eV corresponds to the reference gap reduction for the central C$_{60}$ from the gas phase to this 55-fullerenes cluster configuration  when all fragments are described with their full basis sets.
 }
	\label{fig:fig_55C60}
\end{figure}

Consistently with the results obtained for the first shell of neighbors (Fig.~\ref{fig:fig1}), these calculations confirm that the addition of diffuse orbitals in the test set dramatically helps in reducing the error below the meV with a small number of $N_P$ polarization vectors (compare orange and blue data in Fig.~\ref{fig:fig_55C60}). Further, as compared to the first-nearest neighbors case, a smaller number $N_P$ of polarization vectors is needed to go below the meV error when the constraint on the dipolar polarizability ($l_{max}   = 1$) is imposed. This is the signature that in the long-range, the dipolar response dominates the screening, or reaction field, potential. The combination of diffuse test orbitals with $\zeta=0.01$ bohr$^{-2}$ with the constraint $l_{max} = 1$ leads to an error of the order of 0.1 meV for $N_P = 60$, namely one polarization vector per atom.  This is a dramatic reduction of the size of the polarization basis needed to describe the susceptibility blocks entering the Dyson equation.

\begin{figure*}[t]
	\includegraphics[width=0.8\linewidth]{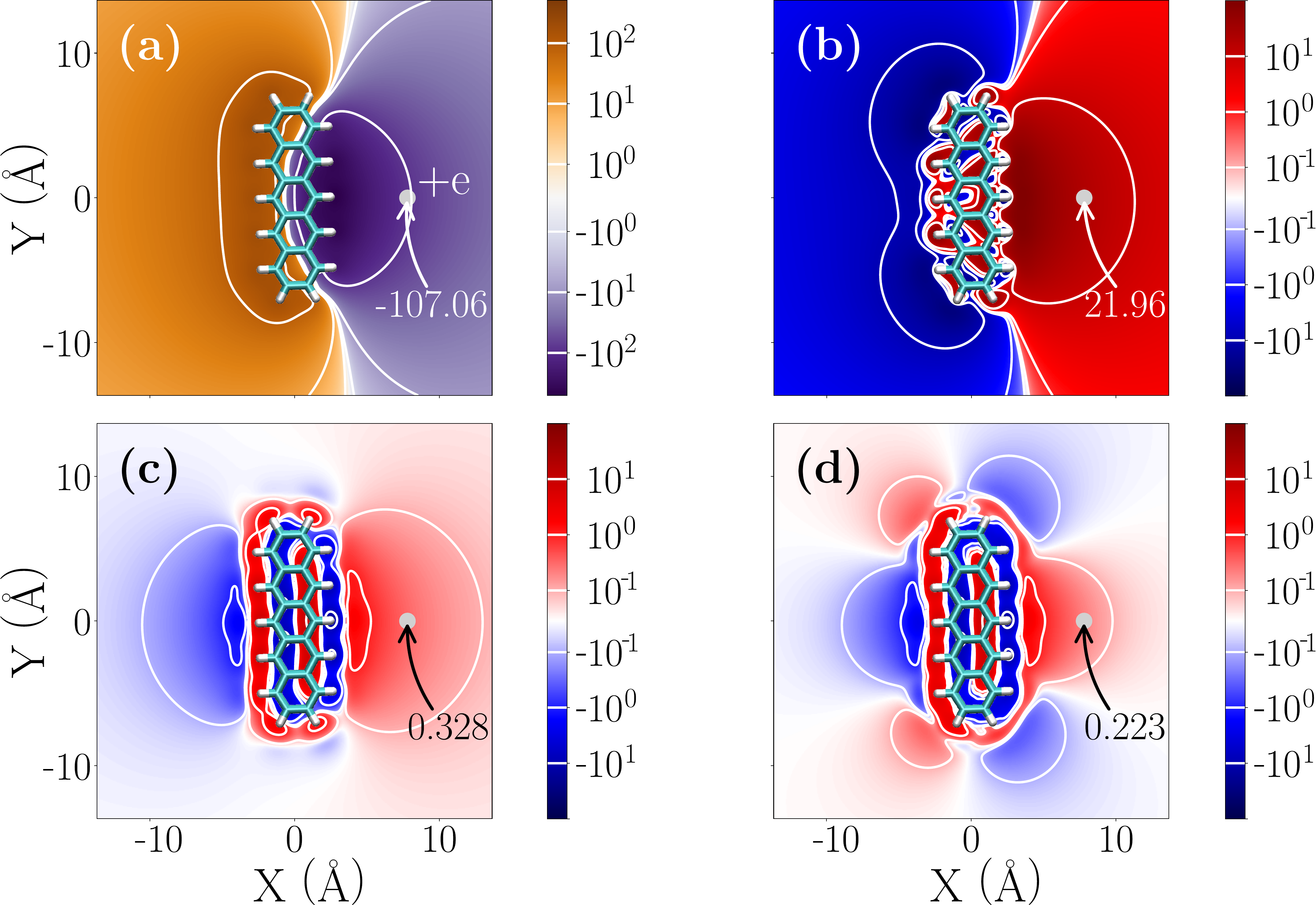}
	\caption{
 In plane $[v\chi v]({\bf r},{\bf r}_0; \omega=0)$ static reaction field generated by a pentacene molecule in response to  a positive unit source charge  (grey dot indicated by  +e) in $\mathbf{r}_0$.
 In (a), the full reaction potential. In (b), (c) and (d) the error $\Delta v_{\text{screen}}(\mathbf{r},\mathbf{r}_0;\omega=0)$ (see Eq.~\ref{eqn:dvreac}) with respect to the full reaction potential upon substituting the full $\chi$ by its low-rank approximation (with 70 polarization vectors). In (b), low-rank $\chi$ model obtained without diffuse functions in the $\lbrace t \rbrace$ test  set; in (c) adding diffuse orbitals to the $\lbrace t \rbrace$ test  set, and in (d) adding further the constraint on the dipolar molecular polarizability. Positions are in angstroms, and units of the reaction field are in meV and in log scale. Equipotentials at $(\pm1,\pm 10, \pm 100)$ meV in figure (a), and at $(\pm0.1,\pm 1, \pm 10)$ meV in figures (b), (c) and (d) are represented by white lines. The value of the reference and of the error associated with the reaction field at the position of the source is indicated. 
 }
	\label{fig:Vreac}
\end{figure*}

As an additional validation, we   plot in Fig.~\ref{fig:Vreac} the screening potential, or reaction field, $v_{\text{screen}}({\bf r},{\bf r}_0)=[v\chi v](\mathbf{r},\mathbf{r}_0)$ 
associated with an elementary positive source point-charge located in ${\bf r}_0$, in the vicinity of a single pentacene molecule. Namely, for a test charge located in ${\bf r}_0$, we plot $v_{\text{screen}}({\bf r},{\bf r}_0)$ as a function of ${\bf r}$ in the pentacene plane. In the case of a single fragment molecule, the reaction field reduces to $v \chi^{(\mathrm{I})} v$, with (I) the index of that molecule. The reference reaction field is provided in Fig.~\ref{fig:Vreac}(a) while the error associated with the model susceptibility, for a fixed $N_P=70$ number of retained polarization vectors among the 2314 vectors of the original def2-TZVP-RI basis, is represented in the other subfigures. Fig.~\ref{fig:Vreac}(b) shows the case  where the test basis set does not contain  diffuse functions. In Fig.~\ref{fig:Vreac}(c) we add diffuse functions ($\zeta =0.01$ bohr$^{-2}$), while Fig.~\ref{fig:Vreac}(d) illustrates the fit of $\chi$ with the same diffuse functions and the constraint $l_{max}=1$.   Clearly, the error associated with the reaction field around a given pentacene molecule (in meV units) is dramatically reduced upon adding diffuse test functions and the constraint on the dipolar polarizability. We purposely replaced the fullerene molecule  by a pentacene to indicate that the accuracy of the present scheme is hardly system dependent.

\subsection{ The C$_{60}$ crystal and surface environments }\label{C60_env}

Beyond the small cluster models, we now study  the evolution of the gap of a fullerene from the gas phase to  a C$_{60}$ face-centered-cubic crystal (fcc) environment. We thus want to calculate the closing  of the gap by screening effects  in the limit of an infinite environment. Such a quantity, labeled $\Delta$Gap below, is also coined the polarization energy. We will consider the cases of bulk C$_{60}$ (fullerite)  and further of a C$_{60}$ at the (111) surface and  sub-surface. Experimental photoemission experiments are very much surface sensitive for organic systems, with a limited penetration depth of the input photons or electrons, so that comparison with the surface location is more appropriate.  The bulk limit is obtained by immersing a C$_{60}$ molecule in a sphere of fullerenes with increasing radius. The surface and subsurface limits are obtained by using half-a-sphere of polarizable fullerenes in the environment. In the absence of wavefunction delocalization (or band dispersion) in the fragment approach, we focus on the peak-to-peak gap, namely the gap between the center of the HOMO and LUMO bands. We emphasize that the absence of permanent ground-state dipole, quadrupole, etc., in fullerene molecules, precludes the influence of any electrostatic crystal field in the ground-state. 

In the present case of fullerene crystals, the reference $\mathbf{X}_g^{(\mathrm{I})}$  susceptibility can be constructed for a single fullerene and the resulting fitted $\widetilde{\mathbf{X}}_g^{(\mathrm{I})}$ matrix can be ``copied'' to form the model susceptibility block associated with each fullerene in the environment. Even though rotational disorder was not explored in this study, rotating the $\widetilde{\mathbf{X}}_g^{(\mathrm{I})}$ matrix, to follow the rotation of a given fullerene, can be easily implemented. Beyond rotations, the effect on the polarization energy of changes in the susceptibility matrix associated with slight atomic distortions around some average equilibrium geometry, is expected to be small but may be explored in future studies. 

The calculations are performed with the parameters described above, namely keeping the full auxiliary  def2-TZVP-RI basis set to describe the susceptibility of the C$_{60}$ of interest for which we perform our embedded $GW$ calculation. The same full auxiliary basis set is used for its first-nearest neighbors.  For the rest of the environment, namely the second-nearest neighbors and those located farther away, we keep $N_P=60$ optimized polarization vectors for each fullerene. 
Diffuse test functions with $\zeta$ = 0.01 bohr$^{-2}$ are adopted with the $l_{max}=1$ constraint. 
We focus on the  gap closing ($\Delta$Gap) from the gas to the dense phase. Such a polarization energy, originating from the screening by the environment, is described at the $\Delta$COHSEX level as emphasized above.

\begin{figure}[htbp!]
	\includegraphics[width=8.6cm]{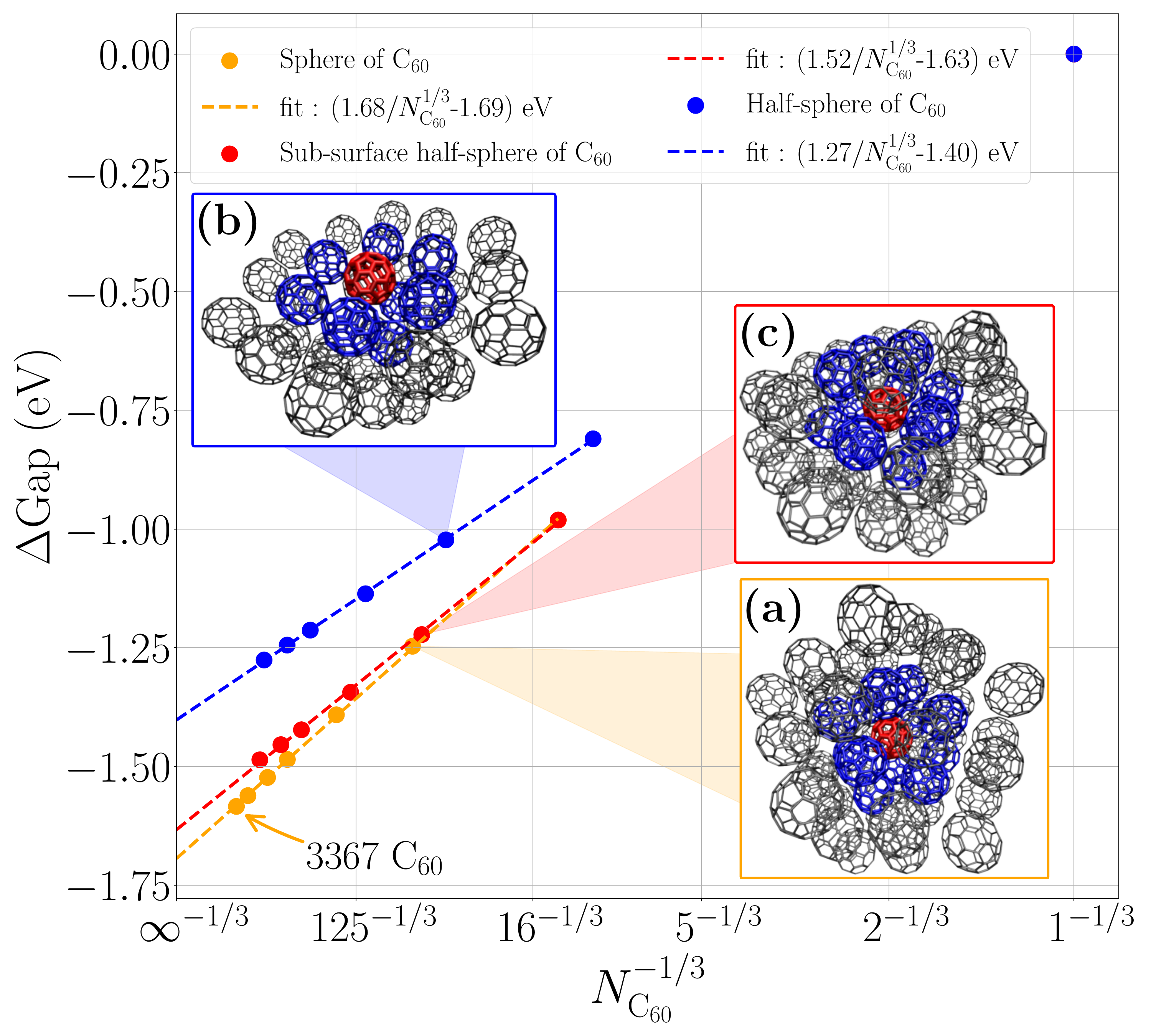}
	\caption{ Evolution of the gap closing $\Delta$Gap, at the static $\Delta$COHSEX level, between a gas phase fullerene and      $N_{\text{C}_{60}}$  systems, as a function of $1/N^{1/3}_{\text{C}_{60}}$. The orange/blue/red   dots show  calculations for one C$_{60}$ embedded at the center of a sphere [Inset (a)], at the surface of a half-sphere [Inset (b)], and at the subsurface of a half-sphere [Inset (c)], respectively. 
 Dashed lines represent     $[a/N^{1/3}_{\text{C}_{60}}+b]$ fits. 
 }
	\label{fig:C60_spheres}
\end{figure}

We plot in Fig.~\ref{fig:C60_spheres} the evolution of the polarization energy, or $\Delta$Gap,  associated with the peak-to-peak gap for a bulk C$_{60}$ (orange dots), a surface C$_{60}$ (blue dots) and a  C$_{60}$ at sub-surface (red dots), as a function of $1/N^{1/3}_{\text{C}_{60}}$, where $N_{\text{C}_{60}}$ represents the number of fullerenes retained in the sphere or half-sphere we use for the environment. 
Analytic derivations show indeed that such a polarization energy converges slowly with a $1/N^{1/3}_{\text{C}_{60}}$ behavior in the asymptotic regime. This asymptotic behavior is confirmed numerically in   Fig.~\ref{fig:C60_spheres} by the straight dashed-line fits, one for each type of system, going through the calculated energies in the large  $N_{\text{C}_{60}}$ limit. 

In the asymptotic infinite bulk size limit, the gap of the central C$_{60}$ is closing by 1.69 eV (orange dashed line) with respect to the gas phase. This can be compared to the polarization energy of the biggest studied system. Our calculations are performed for spheres containing up to 3367 C$_{60}$, representing 202 020 carbon atoms. The gap of such a system closes by 1.58 eV, which represents a difference of 0.11 eV  with respect to the extrapolated infinite size value. This highlights the difficulty to capture the polarization energy with an accuracy within the  0.1 eV threshold when limited size environments are considered. 

In order to compare to the $G_0W_0$@LDA 3.0 eV peak-to-peak gap in a fully periodic bulk calculation performed by Shirley and Louie  in their pioneering study, \cite{Shirley1993} we compute the def2-TZVP $G_0W_0$@LDA gap for an isolated  fullerene. Subtracting the 1.69 eV polarization energy in the bulk limit to our gas phase 4.46 eV $G_0W_0$@LDA HOMO-LUMO gap,
we end up with a 2.8 eV peak-to-peak gap, in fair agreement with the periodic $G_0W_0$@LDA value.  

We also compute  the asymptotic infinite   limit for the gap of one C$_{60}$ located at the (111) surface (see  Inset  Fig.~\ref{fig:C60_spheres}(b)). The blue dashed line  gives an asymptotic closing of the gap amounting to 1.40 eV, which is in good agreement with the experimental values of 1.1 eV,\cite{Reihl1994} 1.2 eV \cite{Weaver1992,Benning1992} or 1.4 eV. \cite{Lof1992,Takahashi1992}  Such values were obtained by taking the provided experimental peak-to-peak gaps subtracted to the experimental 4.9 eV gap value for a fullerene in gas phase. \footnote{  See the NIST website: \url{https://webbook.nist.gov/chemistry/}  }
Alternatively, taking our def2-TZVP  ev$GW$@PBE0 5.1 eV gap value for C$_{60}$ in the gas phase, and adding the calculated polarization energy at the surface, one obtains a surface peak-to-peak gap of 3.7 eV, within the 3.5-3.8 eV experimental range. The presence of a metallic substrate when performing photo-emission experiments, potentially enhancing the screening in the limit of few C$_{60}$ layers, and alternatively the limited screening from the fullerene crystal in the few layer limits, may explain variations between experimental values. On the theoretical side, the influence of the fragment approximation, together with treating screening effects in the static COHSEX limit, remains to be studied.

We further compute  the polarization energy for one C$_{60}$ at the sub-surface (see Inset  Fig.~\ref{fig:C60_spheres}(c)). The red dashed fit provides an asymptotic infinite size polarization energy of 1.63 eV. This value, closer to the bulk limit (difference of 0.06 eV) than the surface limit (difference of 0.23 eV), tends to show a rapid convergence of the polarization energy with respect to the depth of the considered fullerene. Namely, a fullerene in the subsurface presents properties already close to bulk case. 


\begin{figure}[htbp!]
	\includegraphics[width=8.6cm]{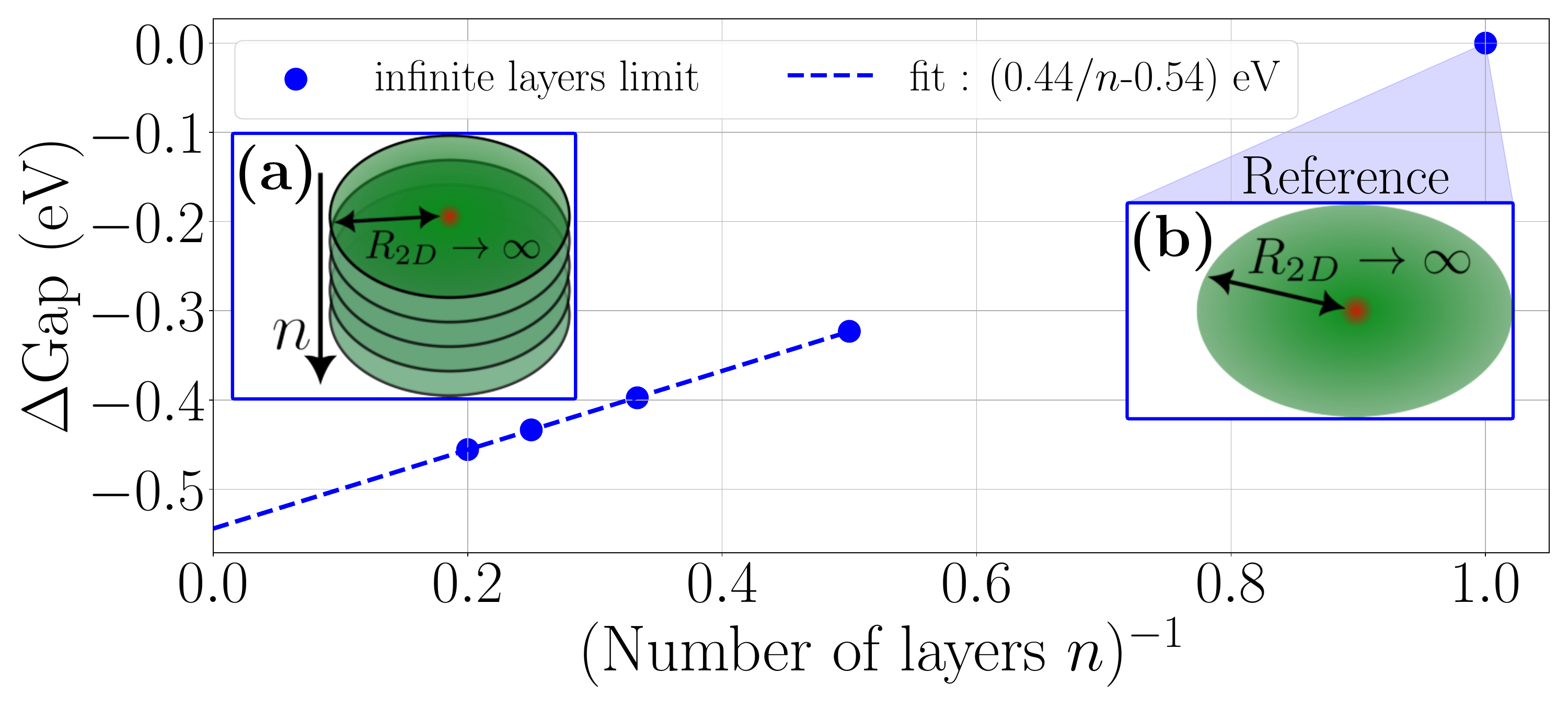}
	\caption{Gap evolution   for one surface C$_{60}$ [in red, Inset (a)], as a function of the inverse   number $n$ of C$_{60}$ layers. Here the  reference is the gap of a fullerene in an infinite C$_{60}$ monolayer   [see Inset (b)]. The infinite size layer(s) limit is  obtained by extrapolating disk(s) to their infinite $R_{2D}$ radius.  
 The dashed line  represents the  $[a/n+b]$ fit.
 }
	\label{fig:C60_plans}
\end{figure}

We finally conclude this study by making a connection to periodic slab calculations, namely the traditional approach for the study of surfaces with periodic boundary conditions. 
Along that line, we converge the polarization energy through the addition of C$_{60}$ infinite sublayers, rather than by increasing the radius $R$ of a  half-sphere of polarizable molecules. Such a representation can be obtained by creating stacks of disks with increasing lateral radius $R_{2D}$, as represented in the Inset   Fig.~\ref{fig:C60_plans}(a), extrapolating the polarization energy to infinity for a given number ($n$) of layers with an asymptotic scaling low in $1/R_{2D}^{2}$ (see Ref.~\citenum{Amblard_2022}). 
Taking as a reference an infinite size  monolayer, as illustrated in the Inset   Fig.~\ref{fig:C60_plans}(b), the polarization energy scales as $(1/n)$. We emphasize that for (n=5), only 85$\%$ of the polarization energy is captured. 
Summing the 0.85 eV polarization energy of one C$_{60}$ embedded in an infinite size monolayer, and the asymptotic polarization energy of 0.54 eV coming from an infinite number of layers with respect to a single monolayer, we find a total polarization energy of 1.39 eV. This value is nearly identical to the 1.40 eV value obtained with the ``half-sphere'' surface approach.

\subsection{ Discussion on CPU and memory requirements }

The calculation for the biggest studied system made of 3367 C$_{60}$, amounting to 202 020 atoms as discussed in section \ref{C60_env}, required around 8000 total CPU hours distributed on 720 cores. Alternatively, this represent a typical wall-time (time to completion of the run) of approximately 11 hours. Such a calculation was performed using the Irene SKL partition (Intel Skylake 8168 processors, with a base frequency of 2.70 GHz) of the IRENE supercomputer from GENCI-IDRIS. Only 2.5 terabytes of memory were used to study this system, making possible to run such calculation on  smaller size computer clusters.  Such a small memory footprint was made possible thanks to the small dimension (60) of the polarization basis for each fragment in the environment. Let's stress out here that describing each fullerene by its  full def2-TZVP-RI basis, of dimension 5700, would lead to a memory footprint of 2.6 petabytes for each related matrices, with similar requirements for the total Coulomb potential $\mathbf{V}$ or the total susceptibility matrix $\mathbf{X}$. It would have been impossible to store any of these big matrices on the 310 terabytes available on the supercomputer used for this study. The present numbers are certainly indicative and may change depending on the chosen parameters, but they illustrate how efficient QM/QM (GW/RPA) calculations can be when using compressed susceptibility blocks for the environment.


 \section{Conclusion}

 We have presented a fully \textit{ab initio} QM/QM embedded $GW$ calculation with a polarizable environment containing up to 200 000 atoms, with total typical CPU timings below 10 000 hours at the def2-TZVP level. Such calculations are made possible by adopting a fragment, or sub-system, approximation. Further, the susceptibility matrices associated with the  fragments in the environment are reduced on-the-fly to a very low-rank representation, with block dimensions equivalent to the number of atoms in the fragment. This approach yields to representations even more compact than standard semi-empirical approaches based on polarizable atoms described by  onsite $\lbrace p_x, p_y, p_z \rbrace$ local induced dipoles, namely 1 degree of freedom per site as compared to the 3 degrees of polarizable atoms.  Such a reduction allows  inverting the Dyson equation for the screened Coulomb potential $W$ with very limited CPU and memory requirements.  
 
 The present scheme is certainly far from being optimal. The choice of the same localization parameter ($\zeta$) for all (\textit{s,p,d,f,g}) diffuse channels in the ensemble of test functions can e.g. lead to forthcoming improvements. Our main point was however to show that very simple choices could already dramatically help in constructing model susceptibility operators both extremely compact and accurate in reproducing medium-to-long-range reaction fields. 
 
 As future directions, the scheme described here above can be merged with subsystem-DFT techniques \cite{neugebauer-wires-2014,Ratcliff2020,Dawson2020} used to improve the input Kohn-Sham eigenstates, allowing to exploit   fragment approximations   while accounting for the frozen density of the neighboring molecules. In particular, the electrostatic field generated by the environment in the ground-state, of crucial importance in organic media composed of molecules with a  permanent dipole, quadrupole, etc. \cite{D_Avino_2014,varsano-ctc-2014,poelking-natmat-2015,D_Avino_2016,jingli-horizon-2019} can be further accounted for at the DFT level. Further, the block-diagonal form of the susceptibility can also be improved within a cluster expansion technique, \cite{Fujita_2021} calculating the gas phase susceptibility of pairs  of interacting neighboring fragments, building a susceptibility matrix that is tridiagonal by blocks rather than strictly diagonal.   
 
 Finally, the $GW$ formalism allows to define a dynamical dielectric response, going beyond the standard low-frequency limit in the optical range of common PCM or QM/MM approaches. The importance of such dynamical corrections needs to be assessed, calculating the polarization energies within the full $GW$ formalism rather than its static COHSEX limit.   
 



\begin{acknowledgments} 
 DA is indebted to ENS Paris-Saclay for his PhD fellowship. This work was performed using HPC resources from GENCI-IDRIS (Grant 2021-A0110910016).
 XB and ID acknowledge support from the French Agence Nationale de la Recherche (ANR) under contract ANR-20-CE29-0005.
\end{acknowledgments} 

 \section*{Data Availability Statement}

The data that supports the findings of this study are available within the article.  


\appendix
\section{Computation of the model susceptibility}\label{appendix_coeff}
In this Appendix, we compute the model susceptibility $\widetilde{\mathbf{X}}_g^{(\mathrm{I})}(\omega)$ solution of the equation \ref{eqn:ximod_coeffs}. From now on and for compactness, we dropped the exponent $(\mathrm{I})$ because we focus only on one fragment, and the frequency index $\omega$. Results can be computed separately for each required frequency. 

Keeping the same notations as the ones used in section \ref{Theory}, we denote $\mathbf{B}$ the Coulomb matrix between the auxiliary basis and the test basis, such that its coefficients are $(Q||t)$, where $(\cdot||\cdot)$ indicates a Coulomb integral. We define $\mathbf{\Gamma}$  the Coulomb matrix between the polarization basis and the test basis, with coefficients $(\gamma||t)$, and  $\mathbf{E}$ (respectively $\mathbf{R}$) the overlap matrix between the auxiliary basis set (respectively the polarization basis) and the constraint basis, with coefficients  $\braket{Q}{x^m y^n z^p}$ (respectively $\braket{\gamma}{x^m y^n z^p}$). The equation \ref{eqn:ximod_coeffs} can be rewritten 
\begin{equation}\label{eqn:appendix_coeff}
    \widetilde{\mathbf{X}}_g = \argmin_{\widetilde{\mathbf{X}}} 
\bigg|\bigg|   \mathbf{B}^{\Dag} \mathbf{X}_{g} \mathbf{B}-\mathbf{\Gamma}^{\Dag}\widetilde{\mathbf{X}}\,\mathbf{\Gamma} \bigg|\bigg|^2,  
\end{equation}
with $||\cdot||$ the Frobenius norm, under constraints like the equation \ref{eqn:lagrange}, which can also be rewritten
\begin{equation}\label{eqn:appendix_cstr}
    \mathbf{R}^{\Dag} \widetilde{\mathbf{X}} \mathbf{R}=\mathbf{E}^{\Dag} \mathbf{X}_{g} \mathbf{E}.
\end{equation}

To find this $N\times N$ matrix $\widetilde{\mathbf{X}}_{g}$, we build feasible solutions, namely matrices which satisfy all constraints enforced by the equation \ref{eqn:appendix_cstr}. Then, among all such feasible solutions, we compute the optimal one, which minimizes Eq.~\ref{eqn:appendix_coeff}.

For compactness, we define $\mathbf{O}=\mathbf{E}^{\Dag} \mathbf{X}_{g} \mathbf{E}$ and $\mathbf{Z}=\mathbf{B}^{\Dag} \mathbf{X}_{g} \mathbf{B}$. Writing the compact singular value decomposition (SVD) of $\mathbf{R}=\mathbf{U}^{}_{\mathbf{R}}\mathbf{\Sigma}^{}_{\mathbf{R}}\mathbf{V}^{\Dag}_{\mathbf{R}}$, and assuming that the rank of $\mathbf{R}$ is equal to the number of constraints $N_{\text{cstr}}$, we search for solutions such that
\begin{equation*}
    \mathbf{U}^{\Dag}_{\mathbf{R}}
    \widetilde{\mathbf{X}}
    \mathbf{U}^{}_{\mathbf{R}}=
    \mathbf{\Sigma}^{-1}_{\mathbf{R}}
    \mathbf{V}^{\Dag}_{\mathbf{R}}
    \mathbf{O}
    \mathbf{V}^{}_{\mathbf{R}}
    \mathbf{\Sigma}^{-1}_{\mathbf{R}}.
\end{equation*}
Inverting $\mathbf{U}^{}_\mathbf{R}$ requires to consider the nullspace of $\mathbf{U}_\mathbf{R}^{\Dag}$, of which we write an orthonormal basis as the columns of the $N\times (N-N_{\text{cstr}})$ matrix $\mathbf{K}$ such that $\mathbf{U}_\mathbf{R}^{\Dag} \mathbf{K}= \mathbf{0}$. Under these considerations, $\widetilde{\mathbf{X}}$ has the form
\begin{equation*}
    \widetilde{\mathbf{X}} = 
    \mathbf{U}^{}_{\mathbf{R}}
    \mathbf{\Sigma}^{-1}_{\mathbf{R}}
    \mathbf{V}^{\Dag}_{\mathbf{R}}
    \mathbf{O}
    \mathbf{V}^{}_{\mathbf{R}}
    \mathbf{\Sigma}^{-1}_{\mathbf{R}}
    \mathbf{U}^{\Dag}_{\mathbf{R}}
    +\mathbf{K}\mathbf{M}^{}_{1}+\mathbf{M}^{}_2 \mathbf{K}^{\Dag}.
\end{equation*}
So far, there are some redundancy in the $\mathbf{M}^{}_1$ and $\mathbf{M}^{}_2$ terms which can be lifted by projection on $\mathbf{U}$ and $\mathbf{K}$ supplementary subspaces. At the end, defining $\mathbf{C}= \mathbf{\Sigma}^{-1}_{\mathbf{R}}\mathbf{V}^{\Dag}_{\mathbf{R}}\mathbf{O}\mathbf{V}^{}_{\mathbf{R}}\mathbf{\Sigma}^{-1}_{\mathbf{R}}$, and using the fact that both $\mathbf{O}$ and $\mathbf{Z}$ are symmetric, we find that  
\begin{equation}\label{ansatz}
    \widetilde{\mathbf{X}} = 
    \mathbf{U}^{}_{\mathbf{R}}
    \mathbf{C}
    \mathbf{U}^{\Dag}_{\mathbf{R}}
    +
    \mathbf{K}
    \mathbf{A}^{}_1
    \mathbf{K}^{\Dag}
    +
    \mathbf{U}^{}_{\mathbf{R}}
    \mathbf{A}^{}_2
    \mathbf{K}^{\Dag}
    +
    \mathbf{K}
    \mathbf{A}^{\Dag}_2
    \mathbf{U}^{\Dag}_{\mathbf{R}},    
\end{equation}
where $\mathbf{A}^{}_1$ and $\mathbf{A}^{}_2$ are computed via the equation \ref{eqn:appendix_coeff}.

 To solve this equation, let write  $\pinv{\mathbf{M}}$ the pseudo-inverse of the matrix $\mathbf{M}$. Let $\mathbf{P}$ be the matrix such that $\mathbf{P}=\pinv{\pqty{ \mathbf{K}^{\Dag}\mathbf{\Gamma}}} \mathbf{K}^{\Dag}$, let $\mathbf{I}_N$ be the identity matrix of size $N\times N$ and let $\mathbf{L}$ be the matrix such that $\mathbf{L}=\pinv{ \mathbf{R}}\bqty{\mathbf{I}_N-{\pqty{ \mathbf{\Gamma} \mathbf{P}}}}$. Using the formula of $\widetilde{\mathbf{X}}$ of Eq.~\ref{ansatz} in Eq.~\ref{eqn:appendix_coeff}, and keeping in mind that $\mathbf{U}$ and $\mathbf{K}$ span supplementary subspaces, the derivation condition with respect to $\mathbf{A}^{}_1$ and $\mathbf{A}^{}_2$ leads to
\begin{equation}\label{sol}
    \boxed{\widetilde{\mathbf{X}}_g =   \mathbf{L}^{\Dag} \mathbf{O} \mathbf{L} +\mathbf{P}^{\Dag}  \mathbf{Z} \pinv{ \mathbf{\Gamma}} + (\pinv{ \mathbf{\Gamma}})^{\Dag}  \mathbf{Z} \mathbf{P} - \mathbf{P}^{\Dag} \mathbf{Z} \mathbf{P}}.
\end{equation}
If no constraints is enforced, we have $\mathbf{R}=\mathbf{L}=\mathbf{0}$, $\mathbf{K}=\mathbf{I}_{N}$ and $\mathbf{P}=\pinv{ \mathbf{\Gamma}}$. The formula simplifies thus into
\begin{equation}
   \widetilde{\mathbf{X}}_g = (\pinv{\mathbf{\Gamma}})^{\Dag} \mathbf{Z} \pinv{\mathbf{\Gamma}}.
\end{equation}

\section{Definition of the test basis}\label{annexe:test_basis}
We detail here how the test basis $\{t\}$ is set-up. It is in particular designed to reproduce as best as possible the effects of the susceptibility of reference at short to middle range.
To do so, we rely on test sets that sample uniformly the Coulomb interaction: 
starting from a set of functions $\{t^{}_0\}$, we orthonormalize this set with respect to the Coulomb norm, leading to the test basis $\{t_1\}$. Namely, writing $\mathbf{V}^{}_0$ the Coulomb matrix such that its coefficients are equal to $(t^{}_0||t'_0)$, where $(\cdot||\cdot)$ indicates a Coulomb integral, the $i^{th}$ vector's coordinates of $\{t_1\}$, in the basis set $\{t^{}_0\}$, is given by the $i^{th}$ column of $\mathbf{V}_0^{-1/2}$. This basis set is such that $(t^{}_{1}\|t'_1)=\delta_{t^{}_1,t'_1}$.

We create two of such sets: i) one spanning the auxiliary basis, namely using the notations of the section \ref{Theory}  $\{t^{}_0\}=\{P\}$, and ii)
another test basis $\{t^{d}_1\}$ based on the set of diffuse orbitals $\{t^{d}_0\}$. In this study, we used for $\{t^{d}_0\}$ a set of atom-centered ($\textit{s,p,d,f,g}$)  diffuse Gaussian orbitals with, for sake of simplicity,  the same $e^{- \zeta r^2}$ radial part.

The final test basis $\{t\}$ is the direct sum of $\{t^{}_1\}$ and $\{t^{d}_1\}$. The first set is down-weighted with respect to the second one, here by a factor $1/50$, to strengthen the influence of the diffuse functions during the fitting process.

\section{Optimization of the polarization basis}\label{annexe:pola_basis}
In this Appendix, we explain how we compute the polarization basis $\{\gamma\}$ (see notation of section \ref{Theory}), in which is computed $\widetilde{\mathbf{X}}_g$. More exactly, we develop our method to compute the $\lbrace C_{\gamma P} \rbrace$ coefficient set of Eq.~\ref{extraction_gamma}. Starting from the equation \ref{ansatz}, these coefficients were computed through two sub-problems.

\subsection{Optimization for $N=N_{\text{cstr}}$}
Keeping the same notation as in Appendix \ref{appendix_coeff}, we started by studying the specific problem of $N=N_{\text{cstr}}$, with $\{\gamma_0\}=\{P\}$. To be more explicit, we re-index by $0$ all matrices which depends directly on $\{\gamma_0\}$. $\mathbf{C}^{}_0$ being invertible, with $\rank(\mathbf{C}^{}_0)=N_{\text{cstr}}$, there is a matrix $\mathbf{S}$ such that $\mathbf{A}^{}_1=\mathbf{S}\mathbf{C}^{}_0\mathbf{S}^{\Dag}$, and $\mathbf{A}^{}_2=\mathbf{C}_0\mathbf{S}^{\Dag}$. We search for $\mathbf{S}$ such that
\begin{equation*}
    \mathbf{S}_{\text{opt}}^{(1)}=\argmin_{\mathbf{S}} \left|\left|\mathbf{Z}-\mathbf{\Gamma}_0^{\Dag} (\mathbf{U}^{}_{\mathbf{R}_{0}}+\mathbf{K}^{}_0\mathbf{S})\mathbf{C}^{}_0(\mathbf{U}^{}_{\mathbf{R}_0}+\mathbf{K}^{}_0\mathbf{S})^{\Dag} \mathbf{\Gamma}^{}_0\right|\right|.
\end{equation*}
The solution $\mathbf{S}^{(1)}_{\text{opt}}$ of this non-linear optimization problem is computed by a gradient descent algorithm. 

We keep $\mathbf{U}^{}_\mathbf{R}=\mathbf{U}^{}_{\mathbf{R}_0}+\mathbf{K}^{}_0\mathbf{S}^{(1)}_{\text{opt}}$ as a subset of the researched optimal directions $\{ \gamma\}$. More exactly, the $i^{th}$ column of ${\mathbf{U}}^{}_\mathbf{R}$ corresponds to the coefficients $\lbrace C_{\gamma P} \rbrace$ of the $i^{th}$ vector of the required basis set  $\{ \gamma\}$.

\subsection{Optimization problem for $\mathbf{A}^{}_2=\mathbf{0}$}

To find the other coefficients $\lbrace C_{\gamma P} \rbrace$, we use a greedy strategy. We consider only the first and second terms of the left-hand side of the equation \ref{ansatz}, so namely $\mathbf{A}^{}_2=\mathbf{0}$, and write the factorized form of the general symmetric matrix $\mathbf{A}^{}_1=\mathbf{S} \mathbf{\Delta} \mathbf{S}^{\Dag}$, with $\mathbf{\Delta}$ a diagonal matrix. Independently of $\mathbf{\Delta}$, the required coefficients $\lbrace C_{\gamma P} \rbrace$ defining the optimal directions are only determined by the $\mathbf{S}$ matrix.
Using the $N_{\text{cstr}}$ first directions we have found,  given $\mathbf{Z}_{\text{cstr}} = \mathbf{Z} - \mathbf{\Gamma}_0^{\Dag} {\mathbf{U}}^{}_{\mathbf{R}} \mathbf{C}^{}_0 {\mathbf{U}}_{\mathbf{R}}^{\Dag} \mathbf{\Gamma}^{}_0$, and $m=\mathrm{max}(0,N-N_{\text{cstr}})$,  we thus search for solutions $\mathbf{S}_{\text{opt}}^{(2)}$ of the low rank sub-problem
\begin{equation*}
    \mathbf{S}_{\text{opt}}^{(2)}=\argmin_{\mathbf{S} ,\mathbf{\Delta}  \text{ / } \rank({\mathbf{S}}) \leq m} \left|\left| \mathbf{Z}_{\text{cstr}} - \mathbf{\Gamma}_0^{\Dag} \mathbf{K}^{}_0 \mathbf{S}\mathbf{\Delta} \mathbf{S}^{\Dag} \mathbf{K}_0^{\Dag} \mathbf{\Gamma}^{}_0\right|\right|.
\end{equation*}
Defining $\mathbf{J}=\mathbf{K}_0^{\Dag} \mathbf{\Gamma}^{}_0$, and using its SVD $\mathbf{J}=\mathbf{U}^{}_\mathbf{J} \mathbf{\Sigma}^{}_\mathbf{J} \mathbf{V}_\mathbf{J}^{\Dag}$, this problem is equivalent to 
\begin{equation}\label{A_2_0}
\mathbf{S}_{\text{opt}}^{(2)}=\argmin_{\mathbf{S} ,\mathbf{\Delta}\text{ / } \rank({\mathbf{S}}) \leq m}\left|\left|\mathbf{V}_\mathbf{J}^{\Dag} \mathbf{Z}_{\text{cstr}} \mathbf{V}^{}_\mathbf{J}- \mathbf{\Sigma}^{}_\mathbf{J} \mathbf{U}_\mathbf{J}^{\Dag} \mathbf{S} \mathbf{\Delta} \mathbf{S}^{\Dag} \mathbf{U}^{}_\mathbf{J} \mathbf{\Sigma}^{}_\mathbf{J}\right|\right|.
\end{equation}
Using the SVD of $\mathbf{V}_\mathbf{J}^{\Dag} \mathbf{Z}_{\text{cstr}} \mathbf{V}^{}_\mathbf{J}=\mathbf{U}^{}_1\mathbf{\Sigma}^{}_1 \mathbf{V}_1^{\Dag}$, and defining $\mathbf{\Sigma}_1^{(m)}$ the diagonal matrix made of the first $m$ singular values of $\mathbf{\Sigma}^{}_1$, the Eckart–Young–Mirsky theorem leads to a solution $\mathbf{S}_{\text{opt}}^{(2)}$ of Eq.~\ref{A_2_0} such that
\begin{equation*}
    \mathbf{S}^{(2)}_{\text{opt}}=\bqty{\sqrt{\mathbf{\Sigma}_1^{(m)}}\mathbf{V}_1^{\Dag} \mathbf{\Sigma}_\mathbf{J}^{-1}\mathbf{U}^{\Dag}_\mathbf{J}}^{\Dag}\;,\quad\mathbf{\Delta} = \mathbf{V}_1^{\Dag} \mathbf{U}^{}_1.
\end{equation*}
The required coefficients $\lbrace C_{\gamma P} \rbrace$ are given by the columns of $\mathbf{K}^{}_0\mathbf{S}_{\text{opt}}^{(2)}$.

At the end of the process, the coordinates of the $N$ polarization vectors $\{\gamma \}$ in the auxiliary basis, corresponding to the $\lbrace C_{\gamma P} \rbrace$ coefficients, are given by the columns of $\mathbf{U}^{}_\mathbf{R}$ and of $\mathbf{K}^{}_0\mathbf{S}_{\text{opt}}^{(2)}$. In the specific case where no constraints is enforced, $\mathbf{R}^{}_0$ is not defined, nor $\mathbf{U}^{}_{\mathbf{R}_0}$, and $\mathbf{K}^{}_0$ is the identity matrix. So the basis $\{\gamma \}$ is given by the columns of $\mathbf{S}_{\text{opt}}^{(2)}$, with $\mathbf{Z}_{\text{cstr}}=\mathbf{Z}$ and $\mathbf{J}=\mathbf{\Gamma}^{}_0$.

We emphasize that such a method can be easily adapted to another choice of the first guess $\{\gamma_0 \}$, different of the auxiliary basis.

\bibliography{xavbib.bib}


\end{document}